\documentclass[twocolumn,prl,floatfix]{revtex4}
\usepackage{graphicx}
\usepackage{dcolumn}
\usepackage{amsmath}
\usepackage{amssymb}

\begin{document}




{\bf Reply:}  In our Letter \cite{Wuetal:02} we 
presented a numerically {\em exact}, polynomial-time
quantum computer algorithm for simulating the class of {\em general}
pairing Hamiltonians (with arbitrary coupling constants $V_{ml}^r$)
\begin{equation*}
H_{\mathrm{p}}=\sum_{m=1}^{N}\frac{1}{2}\varepsilon _{m}\sigma
_{m}^{z}+\sum_{r=\pm }\sum_{l>m=1}^{N}\frac{1}{2}V_{ml}^{r}(\sigma
_{m}^{x}\sigma _{l}^{x}+r\sigma _{m}^{y}\sigma _{l}^{y}),
\label{eq:Hp}
\end{equation*}
[see Eq.~(1) in \cite{Wuetal:02} for definitions]. 
The Comment by Dukelsky {\it et al.} \cite{Dukelsky:comment} 
reviews an interesting but numerically {\em approximate} 
classical algorithm [density matrix
renormalization group (DMRG)] which they claim to be applicable for
the same problem. It is this difference
between numerically approximate and exact algorithms that essentially
renders irrelevant the criticism of our Letter \cite{Wuetal:02} in the
Comment \cite{Dukelsky:comment}.
There is no known classical algorithm for simulating
the class of general pairing models 
that is both efficient (polynomial) {\em and} numerically exact. There is
such a quantum algorithm: the one proposed in our Letter.

Unfortunately, there is a misleading and confusing statement in the
Comment \cite{Dukelsky:comment}: Dukelsky {\it et al.} write in the
third paragraph that (all
emphasis ours) ``It was shown in this
context that only by resorting to the {\em exact} numerical
solution [4], using the DMRG ...''. However, the DMRG is {\em not} an exact
algorithm. It is an approximation, as in fact stated by the 
authors in their Comment a few lines after their claim 
that the DMRG is exact. Indeed, citing Dukelsky and
Sierra's own paper \cite{Dukelsky:99} 
(their Ref.~[4], Ref.~[14] in our Letter): ``We show
that the DMRG gives an accurate {\em approximation} to the exact
ground state if the block is taken to be
the set of particles while the environment is taken to be the set of
holes.'' And on p.174 of the same paper they write: ``The DMRG is a
{\em variational} method and in the region under study we expect our results
to coincide with the exact ones with a relative {\em error} 
less than $10^{-4}$.''

These statements by the authors of the Comment clearly demonstrate that the 
DMRG method is not a numerically exact 
solution of the pairing Hamiltonians. Hence the term ``exact'' in the
the above quote from their Comment is not correct, and the basis for
their criticism of our Letter is invalidated.

The authors have provided evidence that the DMRG can give a good
approximation to the low-lying energy spectrum of 
the {\em reduced} BCS Hamiltonian (with constant pairing interaction
$V_{ml}^{r} \equiv V^{r}$), and can do so in $O(N)$ steps 
\cite{Dukelsky:99}. We 
do not dispute this, and made no statement to the contrary in our 
Letter, since we were only concerned with a numerically exact algorithm. 
However, in spite of their claim to the contrary
in their Comment (``...the DMRG can easily accomadate arbitrary 
pairing matrix elements.''), there is presently no evidence to support the hope
that the quality of the approximation and the efficiency of the
algorithm will remain as good for {\em general} pairing
Hamiltonians (i.e., $V_{ml}^r\neq V^r$ for all $m,l$). 
The burden of proof is upon the authors of the
Comment. Moreover, even if they could demonstrate this, the basic
point remains that theirs is a numerically approximate 
algorithm while ours is exact. Indeed, as we stated in 
our Letter, for half-filled states the dimension of the
Hilbert space grows as $N!/[(N/2)!]^2$ (for even $N$), 
which is super-exponential in $N$, and there
is no known way to avoid this divergence in a classical
algorithm that is numerically exact. In contrast, we have given a
quantum computer algorithm that is numerically exact (in the sense that
it exactly diagonalizes the pairing Hamiltonian), requiring only $O(N^4)$
steps, for arbitrary $N$. Any approximate method, 
including DMRG and projected BCS \cite{Dukelsky:99}, must eventually
be checked against numerically exact algorithms if and when we have
quantum computers.  

In the second part of the Comment, the authors consider a pairing
model that includes a monopole interaction [third term in their
  Eq.~(1)], not included in our pairing Hamiltonian, so it does not
apply to our Letter. They point out that if the coupling constants $V_{ij}^1$
and $V_{ij}^2$ (in their notation) satisfy certain relations then the
model is exactly solvable. This is an elegant result of which we
were well aware, but that bears {\it no} relation to our work, which 
deals, of course, with the cases for which no analytical solution is 
possible. However, Dukelsky {\it et al.} write 
\cite{Dukelsky:comment}: ``We believe 
that most of the physical problems can be modeled with a pairing
Hamiltonian within the integrable subset ...'', indicating that the
cases that are not analytically solvable are somehow typically not 
physicallyrelevant. This conclusion is invalidated by their own 
parameter count: the ratio of integrable to non-integrable models is
$[(6N+3)/(2N^2-N)] \rightarrow 0$ as $N \rightarrow \infty$, so the
class of integrable models occupies a negligibly small fraction of
the parameter space in the interesting regime of $N\gg 1$.  

In conclusion, the Comment \cite{Dukelsky:comment} discusses 
approximate classical algorithms and analytically solvable instances 
of the pairing model, which are certainly important, but are not relevant
in the comparison to a general and numerically exact algorithm for 
this problem, such as ours.\\

L.-A. Wu, M.S. Byrd, and D.A. Lidar\\
$\;\;\;\;\;\;\;\;\;\;\;\;\;\;\;${\small Department of Chemistry, 
University of Toronto, Toronto, Ontario M5S 3H6, Canada}\\


\end{document}